\documentclass[twocolumn,showpacs,preprintnumbers,amsmath,aps,amssymb]{revtex4}
\usepackage{graphicx}
\usepackage{dcolumn}
\usepackage{bm}

\begin{document}

\title{On Noncommutative Minisuperspace and the Friedmann equations.}
\author{W. Guzm\'an}
\email{wguzman@fisica.ugto.mx}
\author{M. Sabido}
\email{msabido@fisica.ugto.mx}
\author{J. Socorro}
\email{socorro@fisica.ugto.mx}
\affiliation{Instituto de F\'{\i}sica de la Universidad de 
Guanajuato,\\
A.P. E-143, C.P. 37150, Le\'on, Guanajuato, M\'exico.}%
\begin{abstract}
In this paper we present noncommutative version of scalar field cosmology. We find the noncommutative Friedmann equations as well as the noncommutative Klein-Gordon equation. Interestingly the noncommutative contributions are only present up to second order in the noncommutitive parameter. Finally we conclude that if we want  a noncommutative minisuperspace with a constant noncommutative parameter as viable phenomenological model,  the noncommuative parameter  is very small.
\end{abstract}
\maketitle

\section{Introduction}
The initial interest in noncommutative field theory \cite{nekra} slowly but steadily  permeated in the realm of gravity, from which  several approaches to noncommutative gravity \cite{GarciaCompean:2003pm} were proposed. All of these formulations showed that the end result of a noncommutative theory of gravity, is a highly nonlinear theory  and finding  solutions to the corresponding noncommutaive  field equations is technically very challenging. 
Even though working with a full noncommutative theory of gravity looks like a fruitless ordeal, several attempts where made to understand the effects of noncommuativity on different aspects of the universe. In some cases the effects of noncommutativity on the gravitational degrees of freedom where ignored \cite{bran}, but interesting results where obtained in connection with scalar field cosmologies.

One particularly interesting proposal concerning noncommutative cosmology, was presented in  \cite {GarciaCompean:2001wy}, the authors conjectured from the fact  that the noncommutative deformations modify the commutative fields,   the effects of the  full  noncummutative theory of gravity should be reflected in the minisuperspace variables.
 This was achieved  by introducing the Moyal product of functions in the Wheeler-DeWitt equation, in the same manner as is done in noncommutative quantum mechanics (NCQM). The model analyzed  was the Kantowski-Sach comology and was carried out at the quantum level, the authors show that a new states of the universe can be created as a consequence  of to introduce this kind of deformations in the  quantum phase space, several works followed with this main idea  \cite{ Barbosa:2004kp, Pimentel:2004jv}.

Although the noncommutative deformations of the minisuperspace where originally analyzed at the quantum level by introducing the effective noncommutativity on the minisuperspace, classical noncommutative formulations have been proposed, in  \cite{ Barbosa:2004kp} for example the authors  considered  classical noncommutative relations in the phase space for the Kantowski-Sach cosmological model, and they establish the classical noncommutative equations of motion. For scalar field cosmology  the  classical minisuperspace  is deformed and a scalar field is used as the matter component of the universe.   In \cite{Guzman:2007},  the study is focused in the consequences that the noncommutative  deformation causes  on the slow-roll parameter when an exponential potential is considered, it is found that the noncommutative deformation gives a mechanism that ends inflation. In all previous work based on a noncommutative minisuperspace a very explicit lapse function is used, this makes the comparison with the usual cosmic time a bit cumbersome and the only option is to analyze effects that are independent of the chosen gauge.
The main idea of this classical noncommutativity is based on the assumption that modifying the Poisson brackets of the classical theory gives the noncommutative equations of motion, this is done by hand and is not justified in any way.
The main purpose of this paper is to construct the noncommutative equations for noncommutative cosmology.

 We will work with an FRW universe and the matter content is a scalar field with arbitrary potential. This model has been used to explain several aspects of our universe, like inflation, dark energy and dark matter, the main reason is the flexibility of the scalar fields and the simplicity of their dynamics. Noncommutativity in the minisuperspace, will be introduced by modifying the symplectic structure, using the formalism of Hamiltonian manifolds. Once this is achieved noncommutative equivalents of the Friedmann equations are derived. Interestingly the noncommutative deformations only appear up to second order in the noncommutative parameter, further more,  if we want to consider noncommutative minisuperspace based cosmology as a viable phenomenological model the resulting equations seem to favor  two very restrictive possibilities, an extremely small value for the noncommutative minisuperspace parameter, or a very high degree of fine tuning in the parameters of the scalar field potential.

The paper is organized as follows. In section II a short description of Hamiltonian manifolds is presented,   in section III the noncommutative equations for scalar field cosmology are presented. Finally section IV is devoted for discussion and outlook.

\section{Classical deformation of the phase space}
In order to work with the noncommutative deformations of the minisuperspace, we start by analyzing deformations of classical mechanics, in order to achieve the deformation we need an appropriate formulation of classical dynamics. For our purposes we will use the symplectic  formalism of classical mechanics.

Is well known that a Hamiltonian classical mechanic can be formulated  in a $2n$-dimensional differential manifold $\mathcal{M}$  with a \emph{symplectic structure}. This means that  a differential  2-form $\omega$ which is closed and non-degenerate exists, the pair formed by  $(\mathcal{M}$, $ \omega $) is called a \emph{symplectic manifold}. 
In the Hamiltonian manifold, Hamilton's function ${\cal H}$ satisfies
\begin{equation}
i_{X_{\cal H}} \omega = -d{\cal H}
\end{equation}
where $X_H$ is called Hamiltonian vector field.  Specifying  local coordinates on  $\mathcal{M}$, $x^\mu=\{q^i,p^i\}$, the above condition takes an explicit dependance on the  2-form $\omega$
\begin{equation}
\frac{dx^\mu(t)}{dt}=\omega^{\mu\nu}\frac{\partial {\cal H}}{\partial x^\nu}, \label{eqns1}
\end{equation}
where $\omega^{\mu\nu}$ are the components of  $\omega^{-1}$ in the local coordinates $x^\mu$. 

In the symplectic manifold there is a general expression for the  Poisson brackets between two functions in $\mathcal{M}$ based on Hamiltonian fluxes  $\{ f , g\} =\omega \left( X_f , X_g \right)$, which in local coordinates has the familiar form
\begin{equation}
\{ f , g\} =\frac{\partial f}{\partial x^\mu} \omega^{\mu\nu}\frac{\partial g}{\partial x^\nu},
\label{eq3}
\end{equation} 
it is easy to check that the last equation generates the following commutation relations
$\{x^\mu, x^\nu \} = \omega^{\mu\nu}$. 

If we consider the \emph{canonical symplectic structure} $\omega_c$ defined by $\omega_c= dp^i\wedge dq^i$, where $i=1,\dots, n$, we recover the usual Poisson brackets and equations(\ref{eqns1}) are just  Hamilton's equations of classical mechanics. 

The Darboux's theorem establish that every symplectic structure  can be driven to canonical one by a suitable choice of local coordinates in the neighborhood of any point $x\in \mathcal{M}$. However it is possible to find new effects if we consider a non-canonical symplectic structure, for example a magnetic field can appear considering the appropriate $\omega$ (see \cite{libro}). 

This formalism of classical mechanics gives the mathematical framework to construct the noncommutative deformation of the minisuperspace. Using this formalism we can calculate the deformed Poisson brackets, from which we will determine the corresponding equations motion and the resulting algebra is consistent with NCQM. Being  the deformation constructed in the tangent bundle  $T\mathcal{M}$ instead of the symplectic manifold $\mathcal{M}$  all the original symmetries are left intact. This feature is attractive because the classical symmetries used to construct a commutative theory would be present in the deformed theory.

\section{Noncommutative cosmological  equations}

Cosmology presents an attractive arena for noncommutative models, both in the quantum as well as classical level. One of the features of noncommutative field theories is UV/IR mixing, this effectively mixes short scales with long scales, from this fact one may expect the even if the noncommutativity is present at a really small scale, by this UV/IR the effects might be present at an older time of the universe.
Furthermore the presence of the noncommutativity could be related to a minimal size, this idea is from the analogy with quantum mechanics where and uncertainty relation between the momentum and coordinates is present.

Let us start by introducing  the phase space for a homogeneous and isotropic universe with  Friedmann-Robertson-Walker  metric 
\begin{equation}
ds^2 = -N^2(t) dt^2+ e^{2\alpha}\left[ dr^2 +r^2d\Omega\right],
\end{equation}
here we have considered a flat universe, $a(t)=e^\alpha$ is the scale factor of the universe and $N(t)$ is the lapse function, finally we will use a scalar field $\phi$ as the matter content for the model. 
The phase space  and the Hamiltonian function is obtained from the action
\begin{equation}
 S =\int  dx^4 \sqrt{-g} \left[ R + 
 \frac{1}{2}g^{\mu \nu}\partial_\mu \phi \partial_\nu \phi+ 
 V(\phi)
  \right] \, ,
\label{accion}
\end{equation}
where we have used the units $8\pi G=1$.

The Hamiltonian function is calculated as in classical mechanics and is given by
\begin{equation}
{\cal H}=e^{-3\alpha}\left[ \frac{1}{12}P_\alpha^2 -\frac{1}{2}P_\phi^2-e^{6\alpha}V(\phi )\right] ,
\label{mec_nc32}
\end{equation}
where  $V(\phi)$ is the scalar potential, we also  set $N(t)=1$, this means that we will be using the usual cosmic time.

The phase space coordinates for this model  are given by $\{ \alpha, \phi ; P_\alpha, P_\phi \}$, using Eq.(\ref{mec_nc32}) together with Eq.(\ref{eqns1}) and the canonical 2-form $\omega_c$, we find the equations of motion 
\begin{eqnarray}
\dot{\alpha}=\frac{1}{6}e^{-3\alpha} P_\alpha , &\quad&
\dot{P}_\alpha=6e^{3\alpha}V(\phi), \label{class3} \\
\dot{\phi}=-e^{-3\alpha}P_\phi,&\quad&
\dot{P}_\phi=e^{3\alpha}\frac{dV(\phi)}{d\phi}.\nonumber
\end{eqnarray}
From the equations for $\alpha$ and $\phi$ and the Hamiltonian  we construct  the Friedmann equation
\begin{equation}
3H^2= \frac{1}{2}\dot{\phi} (t) + V(\phi),  \label{friedmann}
\end{equation}
the Klein-Gordon equation follows from the Hamilton's equations for $\phi$ and $P_\phi$
\begin{equation}\label{k-g}
\ddot{\phi}+3H\dot{\phi}= -\frac{d
V(\phi)}{d\phi}.
\end{equation} 

In order to find the effects of noncommutativity on the cosmological equations of motion, we follow the symplectic formalism on the phase space to the FRW cosmology  with the scalar field. Lets first  consider the following 2-form $\omega_{nc}=\omega_c+\theta dp_\alpha\wedge dp_\phi$, evidently if $\theta$ is constant $\omega_{nc}$ is closed and invertible, thus $\omega_{nc}$ and the cosmological phase space define a symplectic manifold. The components of $\omega^{\mu\nu}_{nc}$ are
\begin{equation} 
\omega_{nc}^{\mu\nu}=\left(
\begin{array}{cccc}
0&\theta&1&0 \\
-\theta&0&0&1 \\
-1&0&0&0  \\
0&-1&0&0  \\
\end{array}
\right),  \label{mec_nc31}
\end{equation}
 From Eq.(\ref{eq3}), the   Poisson commutation relations are
\begin{eqnarray}
\{ \alpha , \phi \} = \theta, \quad \{p_\alpha, p_\phi \}=0, \label{mec_nc30}\\
\{ \alpha, p_\alpha \}=1, \quad \{\phi , p_\phi \} =1. \nonumber
\end{eqnarray}
This particular choice of $\omega$ is inspired and consistent with the effective noncommutativity used in quantum cosmology\cite{GarciaCompean:2001wy,Guzman:2007}. This algebra is the basis for the noncommutative cosmology, but the question of the validity of this deformed versions of the dynamics remains. 

In order to establish the validity, we turn to the symplectic formalism, we can trivially see that the 2-form $\omega$ is exact, this together with Darboux's theorem ensures  that the new 2-form $\omega_{nc}$ gives the correct equations of motion. Using the new algebra we easily calculate the deformed equations that govern the dynamics
\begin{eqnarray}
\dot{\alpha}&=&\{ \alpha , {\cal H} \}=\frac{1}{6}e^{-3\alpha}P_\alpha - \theta e^{3\alpha}\frac{dV(\phi)}{d\phi}, \label{mec_nc33} \\
\dot{\phi}&=&\{ \phi , {\cal H} \}=-e^{-3\alpha}P_\phi + 6\theta e^{3\alpha}V(\phi),\nonumber  
\end{eqnarray}
we omitted writing the equations for the momenta and the Hamiltonian, as they remain  unchanged under the noncommutative deformation. In order to arrive to Eq.(\ref{mec_nc33}) we used the following formulas
\begin{equation}
\{ \alpha , f(\alpha, \phi) \}=\theta \frac{\partial f}{\partial \phi}, \quad \{ \phi, f(\alpha, \phi) \}=-\theta \frac{\partial f}{\partial \alpha},
\end{equation}
which are calculated from the noncommutative relations (\ref{mec_nc30}).
Using Eq.(\ref{mec_nc33}) and the hamiltonian we arrived to the \emph{deformed Friedmann's equation}
\begin{equation}
\begin{split}
3H^2= \frac{1}{2}\dot{\phi}^2+V(\phi) - & 6\theta a^3\left[ H\frac{dV}{d\phi}+\dot{\phi}V\right]   \\
 - & 3(\theta a^3)^2\left[\left(\frac{dV}{d\phi}\right)^2-6V^2\right].
\end{split} 
\label{fried_nc}
\end{equation}
The Klein-Gordon equation for this non-canonical 2-form can be calculated from equation (\ref{mec_nc32}) and (\ref{mec_nc33}) giving 
\begin{equation}
\ddot{\phi} + 3H\dot{\phi}=-\frac{dV}{d\phi}+6\theta a^3\left[ \dot{\phi}\frac{dV}{d\phi}+6HV \right].
\label{kg_nc}
\end{equation}
we can also clearly see that in  the limit $\theta \to 0$ we recover the usual equations of scalar field cosmology. 
These are the noncommutative Friedmann equations for scalar field cosmology, this equations are derived for an arbitrary potential to the scalar field. 

\section{Discussion and Outlook}
In this short paper we have constructed a model of noncommuative scalar field cosmology. The basic assumption is  that the dynamics can be constructed from a new closed and non degenerate diferential 2-form $\omega_{nc}$ on the Hamiltonian manifold $(\mathcal{M,\omega})$, constructed from the minisuperspace. This gives a modified Poisson algebra among the minisuperspace variables that is consistent with the assumption taken in noncommutative quantum cosmology\cite{GarciaCompean:2001wy,Barbosa:2004kp,Pimentel:2004jv}.

The modified equations have the correct commutative limit when the noncommutative parameter vanishes. An intriguing feature is that the corrections only appear up to second order in $\theta$, from this observation we can see that even if the noncommutative parameter was large the effective noncommutative equations have rather simple modifications.

Another simplification arises, for the exponential potential, for example the quadratic term on $\theta$ on Eq.(\ref{fried_nc}) can vanish if we take an exponential potential $V(\phi)=V_0e^{-\lambda\phi}$ and choosing $\lambda=\sqrt{6}$, then from a high degree of fine tuning the equations are further simplified. Further more there would be epochs in when the terms in the brackets multiplied by the noncommutative parameter may vanish, giving  dynamics similar to the commutative universe, but again this will only be achieved under very particular conditions on the potential.

To study the effects of noncommutative in dark energy, dark matter or inflation, we only need to  solve equations (\ref{fried_nc}) and (\ref{kg_nc}) for the particular potential that explains each of the aspects mentioned before. Even if the noncommutative terms look simple, analytical solutions to the equations are difficult to find, but a complete analysis can be done numerically.
Unfortunately things are not as simple as that, a closer look on the noncommutative corrections, we see that these are weighted by the  product $\theta a^3$, being these terms proportional to the volume of the universe the noncommutative corrections would dominate the dynamics at late times. It seems that in order to have some plausible evolution, the minisuperspace noncommutative parameter should be very small, of order of the inverse of the current volume of the universe. Taking this in to account  the effects of noncommutativity will almost disappear at the early epochs of the universe and would be  relevant a the current epoch. This might seem awkward, but scale mixing is a feature that appears in noncommutative field theory, so this might be an effect of the UV/IR mixing.

In conclusion, noncommutative versions of the Friedmann equations where constructed and argued that the only way that this equations can be phenomenological sensible is by using very specific and fined tuned potentials or  an extremely small value of $\theta$, rendering noncommutativity irrelevant at very early stages of the universe with its effects appearing at older stages of cosmological evolution. Then in order to believe that minisuperspace noncommutativity with a constant noncommutative parameter is viable phenomenologically we only have one option, that the noncommuative parameter  is almost zero.
This might be and unattractive result, as one would expect that the effects of noncommutativity be present at early times or scales and disappear as we go to a larger universe, this picture can be realized  if the noncommutativity parameter changes in time, research in this directions are being constructed and will be reported elsewhere.

\section*{Acknowledgments}
This work was partially supported by CONACYT grants 47641, 51306, 62253, CONCYTEG grant 07-16-K662-062 A01 and PROMEP grants UGTO-CA-3 and DINPO 38.07.

  \end{document}